\documentstyle[12pt,epsfig,axodraw]{article}
\def\be{\begin{equation}}
\def\ee{\end{equation}}
\def\bc{\begin{center}}
\def\ec{\end{center}}
\def\bea{\begin{eqnarray}}
\def\eea{\end{eqnarray}}

\def\nn{\nonumber}
\def\auxy{{F_y}}
\def\auxz{{F_z}}
\def\cl{{\cal L}}

\def\ev{{\rm \; eV}}
\def\fy{{\psi_y}}
\def\fz{{\psi_z}}
\def\fyb{{\overline{\psi_y}}}
\def\fzb{{\overline{\psi_z}}}
\def\gev{{\rm \; GeV}}

\def\kev{{\rm \; keV}}
\def\ls{{\Lambda_S}}
\def\lyy{{\Lambda_{yy}}}
\def\lyz{{\Lambda_{yz}}}
\def\lzz{{\Lambda_{zz}}}
\def\lyyy{{\Lambda_{yyy}}}
\def\lyyz{{\Lambda_{yyz}}}
\def\lyzz{{\Lambda_{yzz}}}
\def\lzzz{{\Lambda_{zzz}}}
\def\mpl{M_{\rm P}}

\def\ov{\overline}
\def\qbri{{\Lambda_0^2 \over 16 \pi^2}}

\def\simlt{\stackrel{<}{{}_\sim}}
\def\simgt{\stackrel{>}{{}_\sim}}
\def\tev{{\rm \; TeV}}

\def\yb{{\overline{Y}}}
\def\zb{{\overline{Z}}}

\catcode`@=11
\def\marginnote#1{}
\newcount\hour
\newcount\minute
\newtoks\amorpm
\hour=\time\divide\hour by60
\minute=\time{\multiply\hour by60 \global\advance\minute by-\hour}
\edef\standardtime{{\ifnum\hour<12 \global\amorpm={am}%
        \else\global\amorpm={pm}\advance\hour by-12 \fi
        \ifnum\hour=0 \hour=12 \fi
        \number\hour:\ifnum\minute<10 0\fi\number\minute\the\amorpm}}
\edef\militarytime{\number\hour:\ifnum\minute<10 0\fi\number\minute}
\def\draftlabel#1{{\@bsphack\if@filesw {\let\thepage\relax
   \xdef\@gtempa{\write\@auxout{\string
      \newlabel{#1}{{\@currentlabel}{\thepage}}}}}\@gtempa
   \if@nobreak \ifvmode\nobreak\fi\fi\fi\@esphack}
        \gdef\@eqnlabel{#1}}
\def\@eqnlabel{}
\def\@vacuum{}
\def\draftmarginnote#1{\marginpar{\raggedright\scriptsize\tt#1}}
\def\draft{\oddsidemargin 0.0truein
        \def\@oddfoot{\sl preliminary draft \hfil
        \rm\thepage\hfil\sl\today\quad\militarytime}
        \let\@evenfoot\@oddfoot \overfullrule 3pt
        \let\label=\draftlabel
        \let\marginnote=\draftmarginnote
   \def\@eqnnum{(\theequation)\rlap{\kern\marginparsep\tt\@eqnlabel}%
\global\let\@eqnlabel\@vacuum}  }
\catcode`@=12
\begin{document}
\begin{titlepage}
\vspace*{-1cm}
hep-ph/9805282
\hfill{CERN-TH/98-149}
\\
\phantom{bla}
\hfill{DFPD-98/TH/20}
\vskip 1.0cm
\begin{center}
{\Large\bf Four-fermion interactions and sgoldstino masses \\ 
in models with a superlight gravitino\footnote{Work supported in 
part by the European Commission TMR Programme ERBFMRX-CT96-0045.}}
\end{center}
\vskip 0.8  cm
\begin{center}
{\large 
Andrea Brignole\footnote{E-mail address: 
brignole@padova.infn.it}$^,$\footnote{Also at
Departamento de F\'{\i}sica, Instituto Superior
T\'ecnico, P-1096 Lisboa, Portugal},
Ferruccio Feruglio\footnote{E-mail address: 
feruglio@padova.infn.it}$^,$\footnote{Also at 
CERN, Theory Division, CH-1211 Geneva 23}
and 
Fabio Zwirner\footnote{E-mail address: 
zwirner@padova.infn.it}}
\\
\vskip .5cm
Istituto Nazionale di Fisica Nucleare, Sezione di Padova, 
\\
Dipartimento di Fisica `G.~Galilei', Universit\`a di Padova, 
\\
Via Marzolo n.8, I-35131 Padua, Italy
\end{center}
\vskip 1.0cm
\begin{abstract}
\noindent
We discuss the r\^ole of the effective interactions among four
matter fermions in supersymmetric models with a very light 
gravitino. We show that, from a field-theoretical viewpoint, 
no model-independent bound on the gravitino mass can be 
derived from such interactions. Making use of a naturalness 
criterion, however, we are able to derive some interesting 
but not very stringent bounds, complementary to those 
obtained from the direct production of supersymmetric particles.
We also show that, generically, masses for the spin-0 partners
of the goldstino (sgoldstinos) of the order of the gravitino
mass and much smaller than squark and slepton masses do not
obey a naturalness criterion.
\end{abstract}
\vfill{
CERN-TH/98-149
\newline
\noindent
May 1998}
\end{titlepage}
\setcounter{footnote}{0}
\vskip2truecm
{\bf 1.}
In the study of realistic supersymmetric extensions of 
the Standard Model (for reviews and references, see e.g. 
\cite{susy}), the old subject \cite{old,oldsp} of the 
phenomenological implications of a very light gravitino 
was recently revamped in a series of papers 
\cite{bfz3,bfzph,recent,phenorec,newsp}.
 
It is well known that, if the gravitino is light (say, 
$\ev \simlt m_{3/2} \simlt \kev$), then the effective 
interactions of its goldstino components with the fields 
of the Minimal Supersymmetric Standard Model (MSSM) 
play an important phenomenological r\^ole. Pair-production 
of MSSM R-odd particles (sparticles) at colliders is still
controlled by the renormalizable MSSM couplings, but each of
these particles can decay via its effective coupling with
the corresponding ordinary particle and the goldstino. For 
a given sparticle mass, and apart from mixing effects, the
latter coupling is entirely controlled by the gravitino mass 
$m_{3/2}$ or, equivalently, by the supersymmetry-breaking 
scale $F = \sqrt{3} \, m_{3/2} \, \mpl$, where $\mpl \equiv 
(8 \pi G_N)^{-1/2} \simeq 2.4 \times 10^{18} \gev$ is the 
Planck mass~\footnote{We consider here, for simplicity, the 
case of pure $F$-breaking, with $F$ real and positive.}. 

If the gravitino is very light, say $m_{3/2} \ll \ev$, then its 
effective interactions with the MSSM fields are even stronger, 
and additional phenomenological implications must be taken into
account. For example, diagrams involving goldstino exchange 
can be important for the pair-production of MSSM sparticles. 
Also, the gravitino can be produced in association with an 
MSSM sparticle, such as a sfermion or a gaugino. Finally,  
pair-production of gravitinos can be considered, tagged by a 
single photon or a single jet. By combining the phenomenological 
analyses of all these processes, an absolute lower bound on the 
gravitino mass can be established. A first estimate of this bound 
can be obtained \cite{bfzph} by considering the last class of 
processes, in a situation where the MSSM sparticles are sufficiently 
heavy to escape detection. With this method, the present lower 
bound on the gravitino mass can be estimated to be $m_{3/2} 
\simgt 10^{-5} \ev$, corresponding to $\sqrt{F} \simgt G_F^{-1/2} 
\sim 300 \gev$. An important feature of this limit is its 
model-independence, since, apart from some controllable ambiguity 
\cite{bfzeegg}, the goldstino effective interactions in the 
low-energy limit depend only on $m_{3/2} \leftrightarrow \sqrt{F}$.

The case of a very light gravitino is naturally associated with 
the existence of some new dynamics at a scale very close to the 
electroweak one, responsible for the breaking of supersymmetry, 
the generation of supersymmetry-breaking masses for the MSSM 
sparticles and the scalar partners of the goldstino (sgoldstinos), 
and also the non-renormalizable four-fermion effective interactions 
involving four gravitinos, or two gravitinos and two ordinary 
fermions. This unknown dynamics may also generate
effective four-fermion interactions involving ordinary fermions 
only, which are significantly constrained by the Tevatron data 
\cite{tevatron} (we are concerned here with flavour-conserving 
interactions, since the flavour-changing ones can be naturally 
suppressed by suitable flavour symmetries). We may then ask if 
the study of these interactions can lead to indirect, 
model-independent bounds on $m_{3/2} \leftrightarrow \sqrt{F}$, 
comparable with the bounds coming from direct production processes. 
This is the first question that will be addressed in the present paper.

The second question to be addressed here concerns the class of
supersymmetric models \cite{oldsp,newsp} where the sgoldstinos have 
masses much smaller than the MSSM sparticles: we are going to study
the stability of such a situation with respect to quantum 
corrections.

\vspace{1cm}
{\bf 2.}
To keep the discussion as simple as possible, we consider an 
$N=1$ globally supersymmetric model containing only two chiral 
superfields, $Y\equiv(y,\fy,\auxy)$ and $Z \equiv (z,\fz,\auxz)$.
Despite its simplicity, this model should reproduce all the 
relevant aspects of the realistic case: the $Y$ multiplet will 
mimic the r\^ole of the matter superfields of the MSSM (in the
limit of massless quarks and leptons), whereas the $Z$ multiplet 
will contain the goldstino and the (complex) sgoldstino. 
The most general effective Lagrangian with the above field content 
is determined, up to higher-derivative terms, by a superpotential 
$w$ and by a K\"ahler potential $K$. Here we choose:
\be
\label{wtree}
w = \Lambda_S^2 Z \, ,
\ee 
\bea
K & = & Y \yb + Z \zb
- {Y^2 \yb^2 \over 4 \Lambda^2_{yy} }
- {Y \yb Z \zb \over \Lambda^2_{yz} }
- {Z^2 \zb^2 \over 4 \Lambda^2_{zz} }
\nn \\
& & \nn \\
&+ & 
  {Y^3 \yb^3 \over 9 \Lambda^4_{yyy} }
+ {Y^2 \yb^2 Z \zb \over 4 \Lambda^4_{yyz} }
+ {Y \yb Z^2 \zb^2 \over 4 \Lambda^4_{yzz} }
+ {Z^3 \zb^3 \over 9 \Lambda^4_{zzz} }
+ \ldots \, ,
\label{ktree}
\eea
where $(\ls,\lyy,\lyz,\lzz,\lyyy,\lyyz,\lyzz,\lzzz)$ are
all parameters with the dimension of a mass, to be taken
for now as independent, and the dots stand for higher-order
terms in a power-expansion in the $Y$ and $Z$ fields.
Notice that the K\"ahler potential (\ref{ktree}) is
the most general one compatible with a global $U(1)_Y 
\times U(1)_R$ symmetry, preserved by the superpotential
(\ref{wtree}). We recall that the appearance of 
non-canonical terms in $K$ implies that the model
under consideration is an effective theory, valid up to some
energy cutoff $\Lambda_0$ (see the discussion below).
Whilst it is not restrictive to choose $\Lambda_S$ real and 
positive, the signs in front of the higher-dimensional 
operators in $K$ are purely conventional. In the conventions
of eq.~(\ref{ktree}), it is crucial to have positive 
$\Lambda_{zz}^2$ and $\Lambda_{yz}^2$ to obtain a stable 
vacuum, whereas all the remaining parameters in $K$ can have 
either sign.  

It is straightforward to derive the component Lagrangian 
corresponding to the chosen $w$ and $K$. We give here, for 
illustration, some of the lowest-order non-derivative terms.
The expansion of the scalar potential around the origin is
\be
V =  \Lambda_S^4 
+ {\Lambda_S^4 \over \Lambda_{yz}^2} y \ov{y}
+ {\Lambda_S^4 \over \Lambda_{zz}^2} z \ov{z}
+ \ldots =   
F^2 + m_y^2 y \ov{y} + m_z^2 z \ov{z} + \ldots \, ,
\label{vtree}
\ee 
thus $V$ has a local minimum for
\be
\label{vacuum}
\langle y \rangle = \langle z \rangle = 0 \, ,
\;\;\;\;\;
\langle \auxy \rangle = 0 \, ,
\;\;\;
\langle \auxz \rangle = \Lambda^2_S \, .
\ee   
Supersymmetry is spontaneously broken, with vacuum 
energy $\langle V \rangle \equiv F^2 = \Lambda_S^4$, and 
the global symmetry remains unbroken. Notice that the K\"ahler 
metric is canonical at the minimum, so that the fields are 
automatically normalized. The matter sfermion $y$ and the 
sgoldstino $z$ have masses
\be
m_y^2  = {\Lambda_S^4 \over \Lambda^2_{yz} }  \, ,
\;\;\;\;\;
m_z^2  = {\Lambda_S^4 \over \Lambda^2_{zz} }  \, .
\label{masses}
\ee
Notice that the two masses are controlled by two independent
parameters. In particular, a hierarchical relation between 
them could be arranged, at the classical level, by suitably 
choosing those parameters.
Similarly, the non-derivative part of the Lagrangian 
bilinear in the fermion fields reads
\bea
\cl_{2f} & = &
- {\Lambda_S^2 \over \Lambda_{yz}^2}
\left( \fy \fz \ov{y} + h.c. \right) 
- {1 \over 2} {\Lambda_S^2 \over \Lambda_{zz}^2}
\left( \fz \fz \ov{z} + h.c. \right) + \ldots
\nn \\ & & \nn \\ & = &  
- {m_y^2 \over F}
\left( \fy \fz \ov{y} + h.c. \right) 
- {1 \over 2} {m_z^2 \over F}
\left( \fz \fz \ov{z} + h.c. \right) + \ldots 
\label{yuktree}
\eea 
We remark that there is no fermion mass term, as expected from the
facts that $\fz$ is the goldstino and that a mass for the matter 
fermion $\fy$ would break the global $U(1)_Y$. Finally, the 
effective four-fermion interactions are:
\bea
\cl_{4f} & = & 
- {1 \over 4 \Lambda^2_{zz} } \fz \fz \fzb \, \fzb
- {1 \over \Lambda^2_{yz} } \fy \fz \fyb \, \fzb
- {1 \over 4 \Lambda^2_{yy} } \fy \fy \fyb \, \fyb
+ \ldots
\nn \\ & & \nn \\
& = &
- {m_z^2 \over 4  F^2} \fz \fz \fzb \, \fzb
- {m_y^2 \over    F^2} \fy \fz \fyb \, \fzb
- {1 \over 4 \Lambda^2_{yy} } \fy \fy \fyb \, \fyb
+ \ldots 
\label{4ftree}
\eea
The important fact to notice is that, whilst the coefficients 
of the Yukawa interactions and of the four-fermion interactions 
involving at least two goldstinos can be reexpressed in terms 
of the supersymmetry-breaking scale $F$ and the 
supersymmetry-breaking masses $(m_y^2,m_z^2)$, the coefficient 
of the four-fermion interaction involving only matter fermions 
is controlled by an independent mass parameter, $\lyy$. At the
classical level, then, the possibility of a suppression of the 
latter coefficient with respect to the former ones is perfectly 
consistent. Only the knowledge of the underlying dynamics could 
allow us to say more on the relative size of the different 
mass parameters appearing in eqs.~(\ref{wtree}) and~(\ref{ktree}).

\vspace{1cm}
{\bf 3.}
Even if it is mathematically and phenomenologically consistent 
to assume that $\lyy \gg \lyz,\lzz$, no obvious symmetry seems
to be recovered in the limit $\lyy \rightarrow \infty$.
Similarly, we may consistently assume that $\lzz \gg \lyz$,
corresponding to $m_z^2 \ll m_y^2$, but again no obvious 
symmetry is recovered in the limit $\lzz \rightarrow \infty$.
We may then ask how natural such situations are. To answer this 
question, we shall now compute the most divergent contributions
to the one-loop effective action, and use them to estimate
a naturalness bound on the relative size of the mass scales 
controlling the different physical observables of the model.

Thanks to supersymmetry, quartic divergences are absent, and
the most divergent contribution to the one-loop effective action 
is the quadratically divergent one. We should warn the reader 
that, if the cutoff scale $\Lambda_0$ is not very large, also 
the logarithmically divergent and finite contributions may be 
numerically important. However, our simplifying choice of 
considering only the quadratic divergences will be sufficient 
for a qualitative discussion of the naturalness bounds. The 
quadratically divergent contributions to the one-loop effective 
action are summarized by the following renormalization of the 
K\"ahler potential \cite{grk}
\be
\label{quadren}
\Delta_Q K = 
{\Lambda_0^2 \over 16 \pi ^2} 
\left( \log \det K_{\ov{m} n} \right) \, ,
\ee
where $\Lambda_0$ is an ultraviolet cutoff in momentum space 
and $K_{\ov{m} n}$ is the (field-dependent) K\"ahler metric.
Expanding in powers of the fields, we can write the uncorrected
superpotential $w$ and the corrected K\"ahler potential $K_Q = K 
+ \Delta_Q K$ in the same functional form as in eqs.~(\ref{wtree}) 
and (\ref{ktree}),
\be
\label{wren}
w = \hat{\Lambda}_S^2 \hat{Z} \, ,
\ee 
\be
K_Q = \hat{Y} \hat{\yb} + \hat{Z} \hat{\zb}
- {\hat{Y}^2 \hat{\yb}^2 \over 4 \hat{\Lambda}^2_{yy} }
- {\hat{Y} \hat{\yb} \hat{Z} \hat{\zb} \over \hat{\Lambda}^2_{yz} }
- {\hat{Z}^2 \hat{\zb}^2 \over 4 \hat{\Lambda}^2_{zz} } + \ldots \, ,
\label{kren}
\ee
in terms of renormalized fields and parameters\footnote{
Since we have shown the expansion of $K$ up to the sixth 
order in the fields, for consistency we have shown the 
one of $K_Q$ up to the fourth order.}
\bea
\hat{Y} & = & \left[ 1 - {1 \over 2} \qbri
\left( {1 \over \Lambda_{yy}^2} + {1 \over 
\Lambda_{yz}^2} \right) \right] Y \, ,
\label{reny} 
\\ & & \nn \\
\hat{Z} & = & \left[ 1 - {1 \over 2} \qbri
\left( {1 \over \Lambda_{yz}^2} + {1 \over 
\Lambda_{zz}^2} \right) \right] Z \, ,
\label{renz}
\\ & & \nn \\
\hat{\Lambda}_S^2 & = & \left[ 1 + {1 \over 2} 
\qbri \left( {1 \over \Lambda_{yz}^2} + {1 \over 
\Lambda_{zz}^2} \right) \right] \Lambda_S^2 \, ,
\label{renls}
\\ & & \nn \\
{1 \over \hat{\Lambda}_{yy}^2} & = & 
{1 \over \Lambda_{yy}^2} + \qbri \left( 
{4 \over \Lambda_{yy}^4} + 
{2 \over \Lambda_{yy}^2 \Lambda_{yz}^2} +
{2 \over \Lambda_{yz}^4} -
{4 \over \Lambda_{yyy}^4} -
{1 \over \Lambda_{yyz}^4} \right) \, ,
\label{renlyy}
\\ & & \nn \\
{1 \over \hat{\Lambda}_{yz}^2} & = & 
{1 \over \Lambda_{yz}^2} + \qbri \left( 
{3 \over \Lambda_{yz}^4} + 
{2 \over \Lambda_{yy}^2 \Lambda_{yz}^2} +
{2 \over \Lambda_{yz}^2 \Lambda_{zz}^2} -
{1 \over \Lambda_{yyz}^4} -
{1 \over \Lambda_{yzz}^4} \right) \, ,
\label{renlyz}
\\ & & \nn \\
{1 \over \hat{\Lambda}_{zz}^2} & = & 
{1 \over \Lambda_{zz}^2} + \qbri \left( 
{4 \over \Lambda_{zz}^4} + 
{2 \over \Lambda_{zz}^2 \Lambda_{yz}^2} +
{2 \over \Lambda_{yz}^4} -
{4 \over \Lambda_{zzz}^4} -
{1 \over \Lambda_{yzz}^4} \right) \, , \ldots
\label{renlzz}
\eea

The previous results, obtained from the general formula
of eq.~(\ref{quadren}), have a simple diagrammatic 
interpretation. We consider here, for illustration, the 
effective interaction involving four matter fermions, whose
quadratic renormalization is given in eq.~(\ref{renlyy}).
The (component-field) one-loop diagrams contributing to  
eq.~(\ref{renlyy}) are shown in Fig.~1, where the dots denote 
crossed diagrams in (a) and (b), and diagrams with self-energy
insertions on different lines in (d). The contribution
proportional to $1/\Lambda_{yy}^4$ comes from the
$\psi_y$-loops in (a) and the $y$-loops in (b), (c), 
(d); the one proportional to $1/(\Lambda_{yy}^2 
\Lambda_{yz}^2)$ from the $z$-loops in (d); the one
proportional to $1/\Lambda_{yz}^4$ from the $\psi_z$-loop
in (a) and the $z$-loop in (b); the one proportional 
to $1/\Lambda_{yyy}^4$ from the $y$-loop in (c);
the one proportional to $1/\Lambda_{yyz}^4$ from the
$z$-loop in (c). 
\begin{figure}[htbp]
\begin{center}
\begin{picture}(450,100)(0,0)
\ArrowArc(100,50)(25,0,180)
\ArrowArc(100,50)(25,180,360)
\ArrowLine(50,75)(75,50)
\ArrowLine(75,50)(50,25)
\ArrowLine(125,50)(150,75)
\ArrowLine(150,25)(125,50)
\Text(45,25)[r]{$\psi_y$}
\Text(45,75)[r]{$\psi_y$}
\Text(155,25)[l]{$\psi_y$}
\Text(155,75)[l]{$\psi_y$}
\Text(100,80)[b]{$\psi_y$ ($\psi_z$)}
\Text(100,20)[t]{$\psi_y$ ($\psi_z$)}
\Text(155,50)[l]{+ \ldots}
\Text(100,5)[t]{{\bf (a)}}
\DashArrowArc(350,50)(25,0,180){2}
\DashArrowArc(350,50)(25,180,360){2}
\ArrowLine(300,75)(325,50)
\ArrowLine(325,50)(300,25)
\ArrowLine(375,50)(400,75)
\ArrowLine(400,25)(375,50)
\Text(295,75)[r]{$\psi_y$}
\Text(295,25)[r]{$\psi_y$}
\Text(405,75)[l]{$\psi_y$}
\Text(405,25)[l]{$\psi_y$}
\Text(350,80)[b]{$y$ ($z$)}
\Text(350,20)[t]{$y$ ($z$)}
\Text(405,50)[l]{+ \ldots}
\Text(350,5)[t]{{\bf (b)}}
\end{picture}
\end{center}
\begin{center}
\begin{picture}(450,150)(0,0)
\DashArrowArc(100,100)(25,0,360){2}
\ArrowLine(50,75)(100,75)
\ArrowLine(100,75)(150,75)
\ArrowLine(100,75)(75,25)
\ArrowLine(125,25)(100,75)
\Text(45,75)[r]{$\psi_y$}
\Text(70,25)[r]{$\psi_y$}
\Text(155,75)[l]{$\psi_y$}
\Text(130,25)[l]{$\psi_y$}
\Text(100,130)[b]{$y$ ($z$)}
\Text(100,5)[t]{{\bf (c)}}
\DashArrowArc(335,110)(14.14,0,360){2}
\ArrowLine(300,125)(325,100)
\ArrowLine(325,100)(350,75)
\ArrowLine(350,75)(400,125)
\ArrowLine(350,75)(300,25)
\ArrowLine(400,25)(350,75)
\Text(295,125)[r]{$\psi_y$}
\Text(295,25)[r]{$\psi_y$}
\Text(405,125)[l]{$\psi_y$}
\Text(405,25)[l]{$\psi_y$}
\Text(350,120)[bl]{$y$ ($z$)}
\Text(405,75)[l]{+ \ldots}
\Text(350,5)[t]{{\bf (d)}}
\end{picture}
\end{center}
\caption{\em Quadratically divergent diagrams contributing 
to the $\psi_y \psi_y \ov{\psi_y} \, \ov{\psi_y}$ amplitude.}
\label{fig1}
\end{figure}
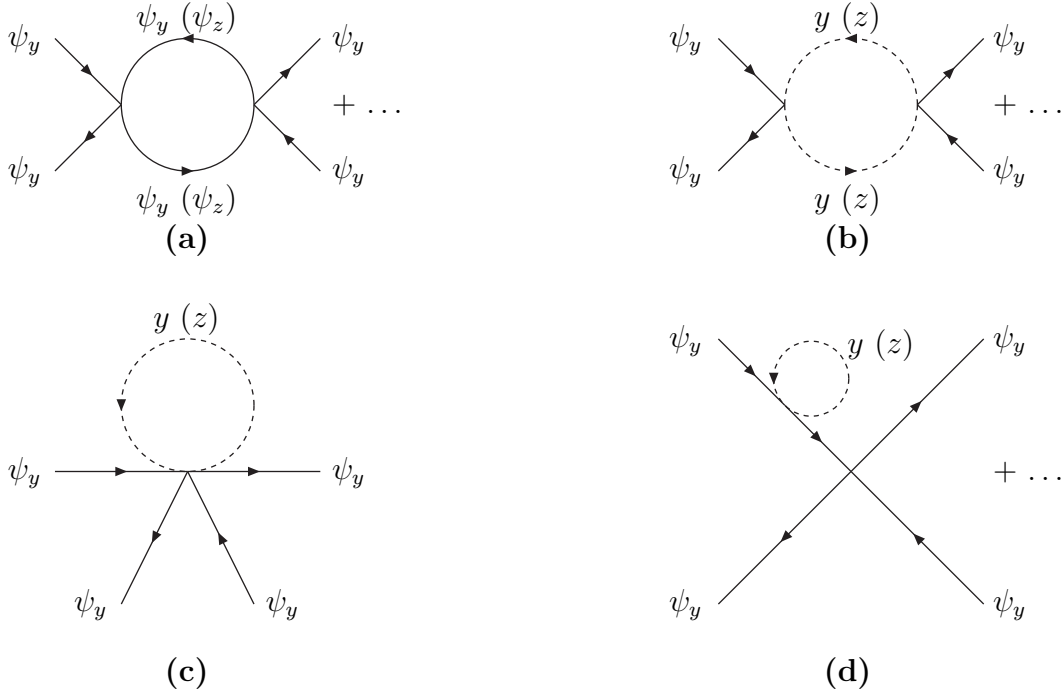
The interaction vertices originate from the couplings of
eq.~(\ref{4ftree}), including those with extra scalars, 
and from derivative couplings involving two fermions 
and two scalars. A similar diagrammatic interpretation
holds for the quadratically divergent corrections to the
other four-fermion interactions and to the scalar masses.
Notice that the renormalized scalar masses can be 
directly obtained from the above formulae as 
${\hat m}_y^2 = {\hat \Lambda}_S^4/{\hat \Lambda}_{yz}^2 $ and 
${\hat m}_z^2 = {\hat \Lambda}_S^4/{\hat \Lambda}_{zz}^2 $. 
We have independently checked this result via explicit 
evaluation of the relevant self-energy diagrams. 

Since all the quadratic divergences can be reabsorbed 
in a redefinition of fields and parameters, all the
predictions obtained from $K_Q$ will be identical in
form to the predictions originally obtained from 
$K$. From the technical point of view, then, a possible
suppression of the four-fermion interactions not 
involving the goldstinos remains viable also at
the quantum level, and the same is true for a possible
suppression of $m_z^2$ with respect to $m_y^2$. On the 
other hand, we may want to take more seriously the physical 
meaning of the cut-off scale $\Lambda_0$, and to ask how much
suppression can be considered natural in the two cases.

In order to proceed, we should first make a statement
about the plausible values that can be assigned to
the cutoff $\Lambda_0$ in the two cases of interest.
We first address the question of four-fermion interactions, 
assuming for simplicity that $\Lambda_{yz} = \Lambda_{zz} \equiv 
\Lambda$, corresponding to $m_y^2 = m_z^2 \equiv m^2 = F^2 /
\Lambda^2$, and that $\Lambda_{yy},\Lambda_{yyy},
\Lambda_{yyz},\Lambda_{yzz},\Lambda_{zzz} \ge 
\Lambda$. Then a fair estimate is
\be
\label{cutoff}
m^2 \simlt \Lambda_0^2 \simlt  16 \pi \Lambda^2 \, ,
\ee
where the lower bound is obvious, and the upper
bound is an estimate of the energy scale at 
which perturbative unitarity is violated by
the most dangerous four-fermion scattering 
amplitudes, proportional to $E^2/\Lambda^2$. 
Incidentally, notice that the interval in
eq.~(\ref{cutoff}) shrinks to a point when
the bound  $m^2 \simlt \sqrt{16 \pi} F$ is 
saturated [the latter bound corresponds to the 
requirement that the spin-0 fields have a particle 
interpretation, $\Gamma(y \to \fy \fz)=2 \Gamma(z 
\to \fz \fz) \simlt m \,$].

We can now see from eqs.~(\ref{renls})--(\ref{cutoff}) that, 
under the previous assumptions, there is no naturalness problem 
for the supersymmetry-breaking scale and for the coefficients 
of the four-fermion amplitudes involving the goldstinos,
since they receive at most relative corrections of order one. 
Instead, if we assume $\Lambda_{yy} \gg \Lambda$ there is a 
potential problem for the four-fermion amplitudes not 
involving the goldstinos, controlled by $\Lambda_{yy}^2$.

To begin with, assume that also the scale parameters
associated with the sixth-order terms of $K$ are 
much larger than $\Lambda$. Then the natural values
of $\Lambda_{yy}$ are those satisfying the bound
\be 
{1 \over \Lambda_{yy}^2} \simgt 
{\Lambda_0^2 \over 8 \pi^2 \Lambda^4} \, .
\label{nat}
\ee
For the two extreme choices of the cutoff scale 
in (\ref{cutoff}), the bound (\ref{nat}) translates
into
\be
{1 \over \Lambda_{yy}^2} \simgt 
{m^6 \over 8 \pi^2 F^4} 
\label{nat1}
\ee
in the least restrictive case, and into 
\be
{1 \over \Lambda_{yy}^2} \simgt 
{2 m^2 \over \pi F^2} 
\label{nat2}
\ee
in the most restrictive one. We shall comment later on 
the phenomenological implications of such inequalities.

Another possibility is that also some of the scale 
parameters associated with the sixth-order terms of 
$K$, in particular $\Lambda_{yyy}$ and $\Lambda_{yyz}$,
are comparable in magnitude with $\Lambda$. Then, due
to the structure of eq.~(\ref{renlyy}), there is the
possibility of cancellations among the different 
contributions. Such cancellations may be accidental,
in which case, beyond a given level of precision, we 
should check the contributions coming from the graphs 
with lower degree of divergence and from higher 
loops. We cannot exclude, however, possible  
cancellations of geometrical nature, related to 
the properties of the K\"ahler manifold. For example,
if the only non-vanishing coefficients in (\ref{ktree})
were those associated with $\Lambda$ and $\Lambda_{yyz}$, 
and the relation $\Lambda^4 = 2 \Lambda^4_{yyz}$ held,
then the correction to $1/\Lambda_{yy}^2$ in (\ref{renlyy}) 
would vanish. More generally, we could look for manifolds
with special properties. The simplest possibility 
that comes to mind is to have an Einstein manifold, 
$R_{\ov{\j} i} = k K_{\ov{\j} i}$, with the hierarchy
$\Lambda_{yy} \gg \Lambda$. If this were possible, the
hierarchy would be automatically stable with respect to 
the correction of eq.~(\ref{quadren}). Unfortunately, it
can be shown that for an Einstein manifold the
relation $\Lambda_{yy} = \Lambda_{zz}$ must hold,
so we should look for more subtle mechanisms.

We will now relax the assumption $\Lambda_{zz} = \Lambda_{yz}$
and see whether a possible hierarchy $\Lambda_{zz} \gg 
\Lambda_{yz}$, corresponding to $m_z^2 \ll m_y^2$,
is stable or not. Assuming that none of the scale parameters 
in $K$ is smaller than $\Lambda_{zy}$, the range 
(\ref{cutoff}) of plausible cutoff values should
now read $m_y^2 \simlt \Lambda_0^2 \simlt 16 \pi \Lambda_{yz}^2$.
Naturalness questions can be addressed by looking again at 
eqs.~(\ref{renls})--(\ref{renlzz}). In particular, we can see 
that assuming $\Lambda_{zz} \gg \Lambda_{yz}$ does not generate 
a naturalness problem for the supersymmetry-breaking scale, 
but does imply a potential problem for the parameter $\Lambda_{zz}$ 
itself. Indeed, eq.~(\ref{renlzz}) shows that, in that case, the 
quantum corrections proportional to $\Lambda_0^2/\Lambda_{yz}^4$ 
can be much larger than the tree-level value $1/\Lambda_{zz}^2$,
especially if we assign to the cutoff $\Lambda_0^2$ the maximum 
(natural) value, of order $\Lambda_{yz}^2$. All this means
that quantum corrections tend to spoil the assumed hierarchy
$m_z^2 \ll m_y^2$ and  drive ${\hat m_z}^2$ close to ${\hat m_y}^2$.  
From this point of view, for example, a situation with sparticle
masses $m_y^2 \gg m_{3/2}^2$ and sgoldstino masses $m_z^2 \simeq 
m_{3/2}^2$ (hierarchy $\Lambda_{yz} \ll \Lambda_{zz} \simeq M_P$)
does not appear natural.

A milder conclusion is reached if we assign to the cutoff $\Lambda_0^2$
the minimum value, i.e. $m_y^2$. Then $m_z^2$ receives quantum corrections
proportional to $m_y^6/F^2$, which do not exceed $m_z^2$ itself provided
$m_z F \simgt m_y^3$. In particular, a situation with $m_z^2 \simeq 
m_{3/2}^2$ would satisfy such a (milder) naturalness criterion 
provided $F^2 \simgt M_P m_y^3$, i.e. $m_{3/2} \simgt m_y^3/M_P$.
Finally, we recall that a tree-level hierarchy $m_z^2 \ll m_y^2$ 
could be maintained at the quantum level also if cancellations among 
different corrections took place in eq.~(\ref{renlzz}), in 
analogy to what observed above when discussing eq.~(\ref{renlyy}).  

\vspace{1cm}
{\bf 4.}
We have shown above that, if we do not invoke any naturalness 
criterion (the most appropriate attitude, in our opinion, 
when discussing {\em model-independent} bounds on $m_{3/2}$ 
and $F$), a suppression of the four-fermion operators
not involving the goldstinos is completely self-consistent.

Nevertheless, it may be instructive to see if, when a 
naturalness criterion is adopted, interesting bounds 
on superlight-gravitino models can be obtained from
the Tevatron bounds on effective four-fermion interactions
involving ordinary fermions. For example, from an analysis
of the dilepton mass spectrum, CDF has published bounds
\cite{tevatron} on possible four-fermion interactions 
involving two quarks and two charged leptons. These bounds
are expressed in terms of a compositeness scale, analogous
(but not identical) to our $\Lambda_{yy}$, and depending on 
the Lorentz and flavour structure of the different operators. 
In the following, we shall denote by $\Lambda_{yy}^*$ the 
putative experimental lower bound on $\Lambda_{yy}$.
When making numerical estimates, we shall use the 
reference value $\Lambda_{yy}^* = 1 \tev$, thus 
taking into account the CDF conventions for the  
normalization of the four-fermion operators. The 
Tevatron experiments should be also sensitive to the 
direct production of sfermion and sgoldstino pairs.
We shall denote by $m^*$ the putative lower bound on
their masses, and use, when making numerical estimates,
the reference value $m^* = 200 \gev$. Combining
the two types of searches, and using eqs.~(\ref{nat1}) 
and (\ref{nat2}), we can derive the corresponding bounds
on the scale of supersymmetry breaking:
\be
\label{loose}
\sqrt{F} \simgt 170 \gev 
\left( {m^* \over 200 \gev} \right)^{3/4}
\left( {\Lambda_{yy}^* \over 1 \tev} \right)^{1/4}
\ee   
for the least restrictive choice of the cutoff scale, and
\be
\label{strict}
\sqrt{F} \simgt 400 \gev 
\left( {m^* \over 200 \gev} \right)^{1/2}
\left( {\Lambda_{yy}^* \over 1 \tev} \right)^{1/2}
\ee
for the most restrictive one. From eqs.~(\ref{loose}) 
and (\ref{strict}) we see that the adoption of 
naturalness criteria on four-fermion (non-goldstino)
interactions leads to bounds on $F$. These bounds are 
comparable with the more direct ones coming from tagged 
gravitino pair-production and from the pair production of 
sfermions and sgoldstinos. To say more, we should 
perform a detailed analysis, taking into account 
the dependences of the different signals on at 
least three independent parameters, e.g.
$(m^2,F,\Lambda_{yy})$. At the level of the toy 
model, this would imply the combined study of 
several processes, such as $\psi_y \psi_y 
\longrightarrow \psi_y \psi_y$, $\psi_y \psi_y 
\longrightarrow \psi_z \psi_z$, $\psi_y \psi_y 
\longrightarrow y y$, $\psi_y \psi_y 
\longrightarrow z z$, \ldots In a fully realistic 
model, there would be additional complications: the 
replacement of the $Y$ superfield with several 
superfields corresponding to left- and right-handed 
quarks and leptons; the introduction of gauge 
interactions, with additional processes and diagrams
involving the gauginos coming into play. However, a 
detailed study of the interplay of the constraints
coming from the different processes goes beyond the 
aim of the present paper.

We conclude by recalling our main results. On the one 
hand, we emphasized that four-fermion interactions not 
involving the goldstinos do not give direct model-independent
bounds on $\sqrt{F}$ or $m_{3/2}$. On the other hand, the
coefficients of such interactions can be indirectly related
to $F$, after considering their renormalization properties 
and adopting some naturalness criterion. The latter viewpoint 
leads to bounds on $\sqrt{F}$ comparable and complementary
to the direct, model-independent bounds. 
As for the sgoldstino mass $m_z^2$ (corresponding to $m_S^2,
m_P^2$ in the more general case considered in the literature),
we have shown that hierarchical situations with $m_z^2 \ll
m_y^2$ (e.g. $m_z^2 \simeq m_{3/2}^2 \ll m_y^2$) are generically 
disfavoured by naturalness considerations, although the possibility 
of cancellations dictated by some symmetry of the underlying fundamental
theory cannot be excluded. 

\vfill{
{\bf Acknowledgements. }
One of us (F.Z.) would like to thank Lisa Randall for 
a chairlift-discussion at the XXXIII Rencontre de 
Moriond that stimulated the present work. We also thank 
Jonathan Bagger, Michelangelo Mangano and Riccardo Rattazzi 
for discussions.
}

\newpage

\end{document}